\documentstyle[preprint,prd,aps,12pt]{revtex}

\author{LUC MARLEAU \footnote{e-mail: lmarleau@phy.ulaval.ca.} and
H. OMARI\thanks{Research supported by the Natural Science and
Engineering Research Council of Canada and by  the  Fonds pour
la Formation de Chercheurs et l'Aide \`a la Recherche du Qu\'ebec.}\\
D\'epartement de Physique, Universit\'e Laval\\
Qu\'ebec, Canada, G1K 7P4}
\title{The Stability of the Weak Skyrmions
}
\date{February 1993
}
\draft
\preprint{LAVAL-PHY-11-93}
\tighten

\begin{document}

\maketitle
\begin{abstract}
We consider a set of gauge invariant terms in higher order effective
Lagrangians of the strongly interacting scalar of the electroweak theory.
The terms are introduced in the framework of the hidden gauge symmetry
formalism. The usual gauge term is known to stabilize the skyrmion but only
in the large vector mass limit. We find that adding higher-order gauge terms
is insufficient to insure stability. We then proceed to analyze other gauge
invariant interaction terms. Some of the conclusions also apply to QCD
skyrmions.
\end{abstract}


\section{Introduction}

Although there has been an enormous amount of theoretical work on the scalar
sector of the electroweak theory \cite{Chanowitz88}, it remains the part of
the Standard Model which is the worst known and this situation will most
probably prevail until we reach accelerator energies that cover scales well
over the electroweak symmetry breaking scale. The minimal description
proposed in the Standard Model relies on spontaneous symmetry breaking but
even in this scheme two possibilities arise: either there is a scalar
particle with mass below 1 TeV (otherwise unitarity is violated in $%
2\longrightarrow 2$ scattering) or the Higgs sector becomes strongly
interacting. In the limiting case where the mass of the Higgs tends to
infinity, the scalar sector (Goldstone particles) becomes equivalent to the
gauged non-linear $\sigma $-model \cite{Veltman77}. The analogy with
low-energy hadron physics suggests the addition of Skyrme-like \cite
{Skyrme61} terms to an effective Lagrangian which in turn could induce new
states, the so-called weak skyrmions. Weak skyrmions are not exclusive to
the spontaneous symmetry breaking mechanism. Indeed, the analogy between QCD
and Technicolor is even more intuitive; in this case the weak skyrmions
could be identified to technibaryons \cite{Farhi81}.

The stabilizing term suggested by Skyrme is added in the context of an
effective theory of scalar fields. (The scalar fields describe pions in QCD
and Goldstone bosons in the Weinberg-Salam theory). Several approaches have
been proposed to stabilize the soliton. The most obvious approach, taken by
Skyrme, is to add just the higher order term in the derivatives of the
fields which is needed for stability. On the other hand, one could consider
a very general effective Lagrangian similar to that proposed by Gipson and
Tze \cite{Gipson81} some time ago and look at the stability properties of
the static solution. Most of the work in that direction is limited to order $%
{\cal O}(\partial ^4)$ terms since it becomes increasingly difficult to keep
track of all possible terms as the number of derivatives increases. Both of
these approaches involve a certain degree of arbitrariness and lack an
underlying principle that would justify the choice of terms. The hidden
gauge symmetry (HGS) formulation proposed by Bando {\it et al. }\cite
{Bando85} is more elegant in that respect. A new gauge symmetry is
introduced. The symmetry is spontaneously broken giving a mass to the new
vector field and it turns out that in the limit where this mass goes to
infinity, the gauge field contribution is equivalent to a stabilizing Skyrme
term \cite{Casalbuoni85}. The stabilization is effective however only in
that limit \cite{Ezawa86}.

Our purpose in this paper is to examine the stability properties of
generalized Lagrangian based on the HGS formalism. We will first consider
higher order gauge terms. We will require that these terms be gauge
invariant in order to retain the spontaneous nature of the gauge symmetry
breaking. They will be generally constructed out of covariant derivatives.
The motivations behind these new objects are the following: (a) The terms
can be viewed as new higher-order gauge terms of some sort and/or as
counterterms when they are dynamically generated.\cite{Bando85} (b) As
mentioned above, the terms are not responsible for symmetry breaking. (c)
Some combinations of gauge terms ({\it e.g. }polynomials of $F_{\mu \nu }$%
's) are especially interesting to many respect. (d) The large vector mass
limit leads to an effective scalar Lagrangian with a Skyrme term. These new
terms will contribute to higher orders in the derivative of the scalar
fields. Previous works \cite{Marleau89-10,Marleau90-13} have shown that
skyrmions exist for certain classes of all-orders effective Lagrangian. (e)
The HGS does not stabilize the soliton in general but that there is a
distinct possibility that the new gauge terms will help. The second part of
our analysis will consider higher-order gauge-invariant interaction terms.

We begin by a description of a strongly-interacting scalar sector of the
Weinberg-Salam theory and of the hidden gauge symmetry formulation. In
section 3, the stability of the weak skyrmion is analyzed with respect to
the HGS. We proceed to prove that the higher-order gauge terms are not
sufficient to stabilize the soliton using arguments similar to that of Ezawa
and Yanagida \cite{Ezawa86}. Section 4 is devoted to other higher-order
gauge-invariant terms, the interaction terms, and contains final remarks.

\section{The scalar sector and hidden gauge symmetry}

The scalar fields of the Standard Model can be written in the form a $%
2\times 2$ matrix $\Phi (x)$:
\begin{equation}
\label{fi}\Phi =\sqrt{2}\left(
\begin{array}{cc}
\phi ^0 & -\phi ^{-*} \\
\phi ^{-} & \phi ^{0*}
\end{array}
\right) .
\end{equation}
In this notation, the Lagrangian of the scalar sector becomes
\begin{equation}
\label{Lscalar}{\cal L}_\Phi {\cal =}\frac 14{\rm Tr\ }D_\mu \Phi D^\mu \Phi
^{\dagger }-\lambda \left[ \frac 12{\rm Tr\ }\Phi \Phi ^{\dagger
}-v^2\right] ^2
\end{equation}
where $v\sim 250$ GeV is the vacuum expectation value (VEV)of the scalar
field , $\lambda $ is related to $v$ through $\lambda =\frac{M_H^2}{8v^2}$
and $D_\mu $ is the covariant derivative corresponding to the $%
SU(2)_L\otimes U(1)_Y$ gauge interactions. When the gauge interactions in (%
\ref{Lscalar}) are switched off (i.e. $D_\mu \rightarrow \partial _\mu $),
the remaining Lagrangian possess a global symmetry under the group $%
G=SU(2)_L\otimes SU(2)_R$. This symmetry is in turn spontaneously broken to
a diagonal subgroup $H=SU(2)_V$ since the scalar field acquire a
non-vanishing VEV, $\left\langle \Phi (x)\right\rangle =v$. In this case, it
is easy to separate the Higgs degree of freedom from the remaining scalar
fields (Goldstone bosons) by rewriting them in terms of a real scalar field $%
\eta (x)$ and a matrix $U(x)\in SU(2)$ respectively such that $\Phi (x)=\eta
(x)U(x)$. In the absence of gauge interactions, ${\cal L}_\Phi $ becomes:
\begin{equation}
\label{LU}{\cal L}_\Phi {\cal =}\frac{\eta ^2}4{\rm Tr\ }\partial _\mu
U\partial ^\mu U^{\dagger }+\frac 14{\rm Tr\ }\partial _\mu \eta \partial
^\mu \eta -\lambda \left( \eta ^2-v^2\right) ^2.
\end{equation}
A large Higgs mass corresponds to a strong coupling $\lambda $ and in the
limit $\lambda \rightarrow \infty $, $\eta $ must be set to the value of the
VEV, $\left\langle \eta \right\rangle =v,$ for consistency. The Higgs field
decouples leaving only Goldstone bosons which take values in the quotient
space $G/H$ and the Lagrangian reduces to that of the non-linear $\sigma $%
-model
\begin{equation}
\label{Lsigma}{\cal L}_\sigma {\cal =}\frac{f^2}4{\rm Tr\ }\partial _\mu
U\partial ^\mu U^{\dagger }
\end{equation}
with $f=v$.

Both the global symmetry group $SU(2)_L\otimes SU(2)_R$ (chiral symmetry)
and the relation of the scalar sector of the theory to the non-linear $%
\sigma $-model suggest that the low-energy effective Lagrangian approach
used in the pion sector of strong interactions can be applied here as well.
The Skyrme Lagrangian is particularly interesting in that respect since it
would induce weak skyrmions constructed as soliton configurations of
Goldstone bosons fields.

In order to introduce a Skyrme-like effective Lagrangian for the scalar
sector of the electroweak theory, we use the formulation of local hidden
gauge symmetry of the non-linear $\sigma $-model \cite{Bando85,Casalbuoni85}%
.The procedure introduce a Yang-Mills field $V_\mu $ and the scalar sector
is represented by the group $G\otimes SU(2)_V$ where $SU(2)_V$ is the gauge
group. The scalar sector can then be conveniently described by two $2\times
2 $ unitary matrices $L(x)\in SU(2)_L$ and $R(x)\in SU(2)_R$ with the
transformation properties%
$$
L(x)\rightarrow h_LL(x)h_V(x),\quad R(x)\rightarrow h_RR(x)h_V(x)
$$
where $h_i\in SU(2)_i$. The most general Lagrangian for $L(x)$ and $R(x)$ up
to order two in derivative of the fields is

\begin{equation}
\label{LR}{\cal L}_S=-{\frac{f^2}4 {\rm Tr\ } }\left[ (L^{\dag }\partial
_\mu L-R^{\dag }\partial _\mu R)^2-a(L^{\dag }\partial _\mu L+R^{\dag
}\partial _\mu R)^2\right] .
\end{equation}
When the $SU(2)_L\otimes U(1)_Y$ electroweak interactions and the new $%
SU(2)_V$ gauge interactions are included, the most general scalar Lagrangian
containing only two derivatives reads:
\begin{equation}
\label{Gauge}{\cal L}_S=-{\frac{f^2}4 {\rm Tr\ } }\left[ (L^{\dag }D_\mu
L-R^{\dag }D_\mu R)^2-a(L^{\dag }D_\mu L+R^{\dag }D_\mu R)^2\right]
\end{equation}
with
\begin{equation}
\label{Dmu}D_\mu L=(\partial _\mu +{\bf W}_\mu )L-L{\bf V}_\mu ,\ \ \ \
D_\mu R=(\partial _\mu +{\bf Y}_\mu )R-R{\bf V}_\mu
\end{equation}
and
\begin{equation}
\label{Wmu}{\bf W}_\mu \equiv i{\frac g2}W_\mu ^k\tau ^k,\quad {\bf Y}_\mu
\equiv i{\frac{g^{\prime }}2}Y_\mu \tau _3,\quad {\bf V}_\mu \equiv i{\frac{%
g_V}2}V_\mu ^k\tau ^k,\ \
\end{equation}
where $g,g^{\prime }$ and $g_V$ are the respective electroweak and hidden
symmetry couplings, $f\sim 250\ GeV$ and $a$ is a positive real number. In $%
{\cal L}_S$ the gauge fields $V_\mu $ are not dynamical. Using the equation
of motion for $V_\mu $
$$
{\bf V}_\mu =\frac 12\left[ L^{\dag }(\partial _\mu +{\bf W}_\mu )L+R^{\dag
}(\partial _\mu +{\bf Y}_\mu )R\right]
$$
leads to the gauged non-linear $\sigma $-model:
\begin{equation}
\label{LG}{\cal L}_{G\sigma }=\frac{f^2}4 {\rm Tr\ } (D_\mu U^{\dag })(D^\mu
U)
\end{equation}
when $U=LR^{\dag }$. Note that for the $U$ fields%
$$
D_\mu U=(\partial _\mu +{\bf W}_\mu )U-U{\bf Y}_\mu
$$
which means that $U$ has no $SU(2)_V$ interactions even though both $L(x)$
and $R(x)$ are interacting with the gauge fields $V_\mu ,${\it i.e.} the
gauge symmetry is hidden.

The gauge bosons become dynamical when we add their kinetic term to the
Lagrangian. They can be interpreted either as new fundamental gauge bosons
or as dynamically generated ones in which case the kinetic term is
introduced as a counterterm with respect to radiative corrections ({\it e.g.
} the $\rho $-meson in QCD, techni-$\rho $ in Technicolor,...). It has the
usual form
\begin{equation}
{\cal L}_X=-{\frac 1{2g_X}} {\rm Tr\ } F_{\mu \nu }^XF^{X\mu \nu }
\end{equation}
where $g_X$ and $F_{\mu \nu }^X$ stand for the coupling and field strength
of each gauge field $X=W,Y,V$. The total Lagrangian is
\begin{equation}
\label{Ltot}{\cal L}_{tot}={\cal L}_{G\sigma }+{\cal L}_W+{\cal L}_Y+{\cal L}%
_V.
\end{equation}
Focusing on the effect of the hidden symmetry contribution, we take the
limit $g_V\gg g,g^{\prime }$ $(g,g^{\prime }\rightarrow 0)$ and neglect
contributions of the gauge bosons $W$ and $Y$ from hereon. The vector bosons
$V$ then receives its mass from the same mechanism as the standard gauge
bosons; in this case we get $m_V^2=f^2g_V^2a$.

\section{Stability of the soliton and higher order gauge terms}

There are no stable solitonic solution for the gauged non-linear $\sigma $%
-model (eq. (\ref{LG})). Furthermore, Ezawa and Yanagida \cite{Ezawa86} have
shown that the hidden gauge symmetry does not stabilize the soliton in
general. On the other hand, it is interesting to note that the total
Lagrangian ${\cal L}_{tot}$ becomes equivalent to the Skyrme Lagrangian in
the limit $m_V\rightarrow \infty ,$ $g,g^{\prime }\rightarrow 0$ where it
induces stable skyrmions. In the more general case where $m_V$ is finite,
the solutions correspond at best to saddle points which are unstable
configurations. Other types of solitonic solutions we may look for are local
minima (classically stable soliton) or global minima (absolutely stable
soliton). Let us first examine how these conclusions are obtained.

The equation of motion for the hidden gauge field ${\bf V}_\mu $ in (\ref
{Ltot}) is ($g,g^{\prime }\rightarrow 0$)
\begin{equation}
\label{Vmu}{\bf V}_\mu ={\frac 12}(L^{\dag }\partial _\mu L+R^{\dag
}\partial _\mu R)+{\frac 1{m_V^2}[}D^{V\nu },[D_{V\mu },D_{V\nu }]]
\end{equation}
with $D^{V\alpha }=\partial _\alpha +{\bf V}_\alpha $. We can use this
equation to perform an expansion in terms of derivatives of the fields \cite
{Marleau92-11}, then
\begin{equation}
\label{Lexp}{\cal L}_{tot}=-{\frac{f^2}4} {\rm Tr\ } L_\mu L^\mu +{\frac
1{32g_V^2}} {\rm Tr\ } [L_\mu ,L_\nu ]^2+{\frac{af^2}{m_V^4}} {\rm Tr\ }
[D^{V\nu },[D_{V\mu },D_{V\nu }]]^2
\end{equation}
where $L_\mu =U^{\dag }\partial _\mu U$. The first and second terms in $%
{\cal L}_S$ are the non-linear $\sigma $-model Lagrangian and the Skyrme
term respectively. Therefore, up to four derivatives the hidden symmetry
approach leads to the Skyrme Lagrangian
\begin{equation}
\label{LSkyrme}{\cal L}_{Skyrme}=-{\frac{f^2}4} {\rm Tr\ } L_\mu L^\mu +{%
\frac 1{32g_V^2}} {\rm Tr\ } [L_\mu ,L_\nu ]^2
\end{equation}
This result which also corresponds to the large $m_V$ limit gives a stable
soliton because of the right sign of the Skyrme term. Another interesting
fact, which is characteristic in this approach, is that the strength of $g_V$
determines the magnitude of the physical parameters of the weak skyrmion
(size, mass, ...) for $m_V\rightarrow \infty $. It also corresponds to a
vector field solution of
\begin{equation}
\label{Vinf}{\bf V}_\mu ={\frac 12}(L^{\dag }\partial _\mu L+R^{\dag
}\partial _\mu R).
\end{equation}

However, it is easy to see that higher-order contributions in the derivative
expansion are necessary and determinant for the stability of the skyrmion
since they regulate the small distance behavior. The mass (or static energy)
of the soliton scales as

\begin{equation}
\label{Mtot}M_{tot}\equiv -\int d^3r\ {\cal L}%
_{tot}=c_2R+c_4R^{-1}+c_6R^{-3}+c_8R^{-5}+...
\end{equation}
when $r$ is scaled according to $r\rightarrow rR^{-1}$ where, $c_n$ is the
coefficient corresponding to the term with $n$ derivatives. The existence of
a global minimum at a finite $r$ ($r\neq 0,\infty $) in the mass indicates
the presence of a stable classical soliton. The large $R$ behavior of the
mass is dominated by the first term and since $c_2$ is positive the size of
the soliton must remain finite. Clearly, in the Skyrme Lagrangian (up to
order four in derivatives), there exists a stable weak skyrmion because $c_4$
turns out to be positive and the series stops at this order. However for $%
{\cal L}_{tot}$, the small distance behavior (large momentum) is dominated
by the remaining terms of the series which are generated by the third term
in (\ref{Lexp}). There is absolute stability if there exist a global minimum
for $M_{tot}(R)$ and if it occurs at a non vanishing skyrmion size R.

For the purpose of showing that the Lagrangian ${\cal L}_{tot}$ does not
induce stable solitons, we introduce the usual hedgehog ansatz for the
chiral field $U$:
\begin{equation}
\label{hedgehog}U=\exp \left( i{\bf \tau }\cdot {\bf \widehat{r}}F(r)\right)
\end{equation}
and the most general spherically symmetric parameterization for the vector
field:

\begin{equation}
\label{Vsph}V_0=0,\quad V_m=i\tau ^n(\delta _{mn}^T\frac{H(r)}{2r}+\widehat{r%
}_m\widehat{r}_n\frac{K(r)}{2r}+\epsilon _{mnp}\widehat{r}^p\frac{G(r)}{2r})
\end{equation}
where $\tau ^n$ are the Pauli matrices, $F(r)$, $G(r)$, $H(r)$ and $K(r)$
are functions of $r$, $\delta _{mn}^T=\delta _{mn}-\widehat{r}_m\widehat{r}%
_n $ and ${\bf \widehat{r}}$ is the unit vector. Both expressions (\ref
{hedgehog}) and (\ref{Vsph}) are invariant under the combined rotations of
spin and isospin. Substituting the last expression in (\ref{Mtot}) leads to
the static energy of the system

$$
M_{tot}=M_{-}+M_{+}+M_V,\quad {\rm with}
$$
$$
M_{-}=\frac{2\pi f^2}{m_V}\int d\rho \ \rho ^2\ \left[ \frac{2\sin {}^2F}{%
\rho ^2}+F^{\prime 2}\right]
$$
\begin{equation}
\label{Mfghk}M_{+}=\frac{2\pi af^2}{m_V}\int d\rho \ \left[ 2(1-\cos
F-G)^2+2H^2+2K^2\right]
\end{equation}
$$
M_V=\frac{2\pi af^2}{m_V}\int d\rho \ \left[ \frac{(G^2-2G+H^2)^2}{\rho ^2}%
+2\left( G^{\prime }-\frac{HK}\rho \right) ^2+2\left( H^{\prime }+\frac{GK-K}%
\rho \right) ^2\right]
$$
where we use the dimensionless variable $\rho =m_Vr$ (recall that $%
m_V^2=f^2g_V^2a$). $M_V$ is the contribution of the gauge field kinetic term
${\cal L}_V$ and $M_{-}$ and $M_{+}$ correspond to the first and second term
of the expression (\ref{Gauge}) for ${\cal L}_S$. By comparison, the Skyrme
Lagrangian involves no gauge field and the static energy is given by

$$
M_{Skyrme}=\frac{2\pi f^2}{m_V}\int d\rho \ \rho ^2\ \left[ \left( \frac{%
2\sin {}^2F}{\rho ^2}+F^{\prime 2}\right) +\frac{\sin {}^2F}{\rho ^2}\left(
\frac{\sin {}^2F}{\rho ^2}+2F^{\prime 2}\right) \right]
$$
The static energy in (\ref{Mfghk}) shows a discrete symmetry under the
transformation $(F,G,H,K)\rightarrow (F,G,-H,-K)$ and clearly the static
solution will be parity conserving only for $H(r)=K(r)=0$ otherwise it will
break parity spontaneously. The $N=1$ topological soliton that we are
interested in corresponds to the boundary conditions $F(0)=\pi $, $G(0)=2$, $%
H(0)=K(0)=0$, and $F(\infty )=G(\infty )=H(\infty )=K(\infty )=0$.

However, if we consider a more specific solution

\begin{equation}
\label{gsin}G(r)=g(r)\sin {}^2\theta ,\quad \ H(r)=g(r)\sin {}\theta \cos
\theta ,\quad K(r)=0\quad
\end{equation}
where $\theta $ is a constant parameter, the static energies reduce to%
$$
M_{+}=\frac{4\pi af^2}{m_V}\int d\rho \ \left[ (1-\cos F)^2+\left(
g^2-2g(1-\cos F)\right) \sin {}^2\theta \right]
$$
$$
M_V=\frac{2\pi af^2}{m_V}\int d\rho \ \left[ \frac{g^2(g-2)^2}{\rho ^2}%
+2g^{\prime 2}\right] \sin {}^2\theta .
$$

For $\theta =\pi /2$, the vector field becomes identical to the monopole
ansatz
\begin{equation}
\label{Vmono}V_0=0,\quad V_m=i\tau \epsilon _{mnp}\widehat{r}^p\frac{G(r)}{2r%
},
\end{equation}
but decreasing $\theta $ from $\pi /2$ to $0$ continously eliminates
smoothly the higher order terms in $M_{tot}$ with the final result
$$
M_{tot}(\theta =0)=\frac{2\pi f^2}{m_V}\int d\rho \ \rho ^2\ \left[ \left(
\frac{2\sin {}^2F}{\rho ^2}+F^{\prime 2}\right) +\frac{2a}{\rho ^2}(1-\cos
F)^2\right]
$$
The global minimum in the static energy (mass of the soliton) is found at
soliton size $R=0$. In other words, a static solution for a gauge field of
the form (\ref{Vsph}) will decay ({\it e.g. }sphaleron) and in the absence
of any restraining term, the soliton will shrink to zero size and disappear.
One may wonder how does this relates to the result obtained for large $m_{V
\text{ }}$ limit ( {\it i.e. }stable skyrmions)? When $m_{V\text{ }%
}\rightarrow \infty $, the vector field hardly propagates and it is forced
into the static configuration (\ref{Vinf}) which has the monopole form (\ref
{Vmono}).

Note that the configuration in (\ref{gsin}) does minimize $M_{tot}$ only for
$\theta =0$ or $\pi /2$. Indeed, minimizing $M_{tot}$ with respect to $K$
leads to the equation of motion%
$$
K=\rho \frac{H(G-1)^{\prime }-\ H^{\prime }(G-1)}{H^2+(G-1)^2+\rho ^2/2}%
=\rho \frac{g^{\prime }\sin {}\theta \cos \theta }{g(g-2)\sin {}^2\theta
+1+\rho ^2/2}.
$$
instead of $K=0$ in (\ref{gsin}). Otherwise the conclusion remains intact.

Consider now a higher-order gauge Lagrangian $\widetilde{\cal L}_V$. We will
be interested in contributions to $\widetilde{\cal L}_V$ which can be
expressed as polynomials of $F_{\mu \nu }^V$'s.
\begin{equation}
\label{Traces} {\rm Tr\ } \left( F_{\mu \nu }^VF^{V\mu \nu }\right) ,\quad
{\rm Tr\ } \left( F_\mu ^{V\nu }F_\nu ^{V\lambda }F_\lambda ^{V\mu }\right)
,\quad {\rm Tr\ } \left( F_{\mu \nu }^VF^{V\mu \nu }\right) ^2,\quad {\rm %
Tr\ } \left( F_\mu ^{V\nu }F_\nu ^{V\lambda }F_\lambda ^{V\rho }F_\rho
^{V\mu }\right) ,\quad \quad {\it etc...}
\end{equation}
These terms are a natural generalization of the gauge field kinetic term
which can be used to build an effective Lagrangian. They are, of course,
manifestly gauge invariant. Our interest here lies mainly in the fact that
in the limit $m_V\rightarrow \infty $, the field strength tends to $F_{\mu
\nu }^V=[D_{V\mu },D_{V\nu }]\rightarrow -\frac 14L^{\dagger }[L_\mu ,L_\nu
]L=-\frac 14L^{\dagger }f_{\mu \nu }L$. Consequently, they also have the
same $SU(2)$ and Lorentz structure as the objects analyzed in refs. \cite
{Marleau89-10} and \cite{Marleau91-11}.

Let us recall some of these results. For the hedgehog static solution one
can write
\begin{equation}
\label{Lheg}L_0=0,\quad L_m=i\tau ^n(\alpha (r)\delta _{mn}^T+\beta (r)
\widehat{r}_m\widehat{r}_n+\gamma (r)\epsilon _{mnp}\widehat{r}^p)
\end{equation}
where $\alpha ,\beta $ and $\gamma $ are
$$
\alpha (r)=\frac{\sin F\cos F}r,\quad \beta (r)=F^{\prime },\quad \gamma
(r)= \frac{\sin {}^2F}r.
$$
The commutator takes the general form

\begin{equation}
\label{fheg}
\begin{array}{ll}
f_{ij}=-2i\tau ^n & [\alpha ^2\epsilon _{ijk}\delta _{kn}^T+(\alpha
^2+\gamma ^2)\epsilon _{ijk}
\widehat{r}_k\widehat{r}_n+\beta \gamma (\delta _{jn}^T\widehat{r}_i-\delta
_{in}^T\widehat{r}_j)+ \\  & \ \alpha (\beta -\alpha )(\epsilon _{ikn}
\widehat{r}_k\widehat{r}_j-\epsilon _{jkn}\widehat{r}_k\widehat{r}_i)].
\end{array}
\end{equation}
The calculations in ref. \cite{Marleau91-11} led to a number of properties
of trace such as those in (\ref{Traces}). First, all traces are found to be
polynomials of the following combinations, $a_{-}=(\alpha ^2+\gamma ^2)$ and
$b_{-}=\beta ^2$. They can be written as
\begin{equation}
\label{fn} {\rm Tr\ } (f_{\mu \nu })^n={\rm const\ \cdot \ }%
\sum_{m=0}^{\left[ \frac n2\right] }\kappa _ma_{-}^{n-m}(b_{-}-a_{-})^m
\end{equation}
where $(f_{\mu \nu })^n$ is a generic form for any combination of $n$ $%
f_{\mu \nu }$'s, $\left[ k\right] $ is the integer part of $\ k$ and $\kappa
_0/\kappa _1$ is found to be $3/n$. For example,%
$$
{\rm Tr\ } f_{\mu \nu }f^{\mu \nu }=16a_{-}[a_{-}+2b_{-}],\quad {\rm Tr\ }
f_{\mu \nu }f^{\nu \lambda }f_\lambda ^{\ \ \mu }=96a_{-}^2b_{-},...\quad
$$
Furthermore, one can construct a special class \cite{Marleau89-10} of such
combinations which is at most linear in $b_{-}$ (or of degree two in
derivatives of $F$). These Lagrangians give a very simple form
\begin{equation}
\label{fn2} {\rm Tr\ } (f_{\mu \nu })^n={\rm const\ \cdot \ }%
a_{-}^{n-1}[3a_{-}+n(b_{-}-a_{-})]
\end{equation}
and lead to a chiral angle equation which is tractable since it is of degree
two, despite the fact that the former involve $2n$ derivatives terms. It
also turns out that for this class of Lagrangians ${\rm const\ }=0$ for $n$
odd $\geq 5$.

Let us now transpose these results to an arbitrary operator $\zeta _\mu $
and the commutator $\zeta _{\mu \nu }\equiv \left[ \zeta _\mu ,\zeta _\nu
\right] $ and consider a Lagrangian from the commutators $\zeta _{\mu \nu
}\equiv \left[ \zeta _\mu ,\zeta _\nu \right] $
\begin{equation}
\label{Traceszeta}\widetilde{\cal L}=a_2 {\rm Tr\ } \left( \zeta _{\mu \nu
}\zeta ^{\mu \nu }\right) +a_3 {\rm Tr\ } \left( \zeta _\mu ^{\ \nu }\zeta
_\nu ^{\ \lambda }\zeta _\lambda ^{\ \mu }\right) +a_4 {\rm Tr\ } \left(
\zeta _{\mu \nu }\zeta ^{\mu \nu }\right) ^2+\ \cdots
\end{equation}
It is easy to see that any $SU(2)$ commutator with the same Lorentz
structure as (\ref{fheg}) may be written as:
\begin{equation}
\label{com}
\begin{array}{ll}
\zeta _{ij}\equiv \left[ \zeta _i,\zeta _j\right] =-2i\tau ^n & [\alpha
_1\epsilon _{ijk}\delta _{kn}^T+\alpha _2\epsilon _{ijk}
\widehat{r}_k\widehat{r}_n+\alpha _3(\delta _{jn}^T\widehat{r}_i-\delta
_{in}^T\widehat{r}_j)+ \\  & \ \alpha _4(\epsilon _{ikn}\widehat{r}_k
\widehat{r}_j-\epsilon _{jkn}\widehat{r}_k\widehat{r}_i)].
\end{array}
\end{equation}
with $\alpha _i\equiv \alpha _i(r)$ and for practical purposes here we set $%
\zeta _0=0$. It is then easy to show that (\ref{fn}) and (\ref{fn2}) apply
to the $\zeta _{\mu \nu }$ commutators by substituting $f_{\mu \nu
}\rightarrow \zeta _{\mu \nu }$ and
$$
\ a_{-}\rightarrow a_\zeta =\alpha _2\ \quad {\rm and\quad }%
a_{-}b_{-}\rightarrow a_\zeta b_\zeta =(\alpha _1+\alpha _4)^2+\alpha _3^2.
$$

In general, a Lagrangian constructed from all orders of $\zeta _{\mu \nu }$
leads to a static energy which can be written as:
\begin{equation}
\label{Mstatic}\widetilde{M}\equiv -\int d^3r\ \widetilde{\cal L}=4\pi \int
r^2dr\sum_{n=1}^\infty \sum_{m=0}^{\left[ \frac n2\right] }\kappa
_{n,m}a_\zeta ^{n-m}(b_\zeta -a_\zeta )^m
\end{equation}
where according to (\ref{fn2}), $\kappa _{n,0}/\kappa _{n,1}=3/n$. The
special class of combinations which is at most linear in $b_\zeta $ gives an
even simpler form since $\kappa _{n,m}=0$ for $m>1$:
\begin{equation}
\label{Xi}\widetilde{M}=4\pi \int r^2dr\left[ 3\chi (a_\zeta )+(b_\zeta
-a_\zeta )\chi ^{\prime }(a_\zeta )\right]
\end{equation}
where $\chi (x)=\sum_{n=1}^\infty \kappa _nx^n$ and $\chi ^{\prime
}(x)=d\chi (x)/dx$. Note that for $\widetilde{M}$ to be positive, it is
sufficient to impose the conditions $\kappa _{n,m}\geq 0$ and $(b_\zeta
-a_\zeta )$ $\geq 0$.

Let us now consider the higher-order gauge terms of (\ref{Traces}). The
field strength in those expressions have the same commutator structure as in
(\ref{com}) and indeed one finds that
\begin{equation}
\label{Dcom}
\begin{array}{ll}
F_{ij}^V=-2i\tau ^n & [\alpha _1\epsilon _{ijk}\delta _{kn}^T+\alpha
_2\epsilon _{ijk}
\widehat{r}_k\widehat{r}_n+\alpha _3(\delta _{jn}^T\widehat{r}_i-\delta
_{in}^T\widehat{r}_j)+ \\  & \ \alpha _4(\epsilon _{ikn}\widehat{r}_k
\widehat{r}_j-\epsilon _{jkn}\widehat{r}_k\widehat{r}_i)].
\end{array}
\end{equation}
with%
$$
\alpha _1=\frac{2G+H^2}{2r^2},\ \alpha _2=\frac{G(G-2)+H^2}{2r^2},\ \alpha
_3=\frac{rH^{\prime }+(G-1)K}{2r^2},\ \alpha _4=\frac{2G-rG^{\prime }-H(H-K)
}{2r^2}\quad
$$
We can then use the above results for the gauge field case by substituting $%
\zeta _{ij}\rightarrow F_{ij}^V$ and $a_\zeta ,$ $b_\zeta \rightarrow a_V,$ $%
b_V$ with
$$
\ a_V=\frac{G^2-2G+H^2}{2r^2}\ \quad {\rm and}
$$
\begin{equation}
\label{avbv}a_Vb_V=\frac 1{4r^4}\left[ \left( rG^{\prime }-HK\right)
^2+\left( rH^{\prime }+(G-1)K\right) ^2\right] .
\end{equation}

In general, an all-order Lagrangian $\widetilde{\cal L}_V$ constructed from
all powers of $F_{\mu \nu }^V$ leads to a static energy which can be
written:
\begin{equation}
\label{MV}\widetilde{M}_V\equiv -\int d^3r\ \widetilde{\cal L}_V=4\pi \int
r^2dr\sum_{n=1}^\infty \sum_{m=0}^{\left[ \frac n2\right] }\kappa
_{n,m}a_V^{n-m}(b_V-a_V)^m
\end{equation}
or, according to (\ref{Xi}), the special class of Lagrangians which leaves
the degree of the chiral angle equation equal to two:
\begin{equation}
\label{XiV}\widetilde{M}_V=4\pi \int r^2dr\left[ 3\chi _V(a_V)+(b_V-a_V)\chi
_V^{\prime }(a_V)\right] .
\end{equation}
It is interesting to note that, even away from the limit $m_V\rightarrow
\infty $, $\widetilde{M}_V$ is only quadratic in the derivatives and this
leads to equations with an overall degree of two. Some examples of stable
solitons were given in ref. \cite{Marleau90-13}; they correspond to the
limit $m_V\rightarrow \infty $ of (\ref{XiV}).

We would like to address here the question of absolute stability and the
presence of a non trivial global minimum in the static energy. In other
words, are the gauge field terms introduced in (\ref{Traces}) sufficient to
stabilize the soliton? We use the same argument as above to prove that this
is not the case. Consider the gauge field configuration in (\ref{gsin}). The
functions $a_V$ and $b_V$ become%
$$
\ a_V=\frac{\sin {}^2\theta }{2r^2}g(g-2)\ ,\quad a_Vb_V=\frac{g^{\prime 2}}{%
4r^2}\sin {}^2\theta .
$$
The static energy
\begin{equation}
\label{Mstatica}\widetilde{M}_V=4\pi \int r^2dr\sum_{n=1}^\infty
\sum_{m=0}^{\left[ \frac n2\right] }\kappa _{n,m}\left[ \frac{\sin
{}^2\theta }{2r^2}\right] ^{n-m}\left[ g(g-2)\right] ^{n-2m}\left[ \frac{%
g^{\prime 2}}2-\frac{\sin {}^2\theta }{2r^2}g^2(g-2)^2\right] ^m
\end{equation}
coming from the gauge terms, falls monotonously as $\theta $ goes from $\pi
/2$ to $0.$ Therefore, all gauge terms contributions vanish for $\theta =0$
leaving no stabilizing terms in $M_{tot}$. Given a soliton configuration, it
will tend to shrink to zero size and vanish. For this kind of generalized
Lagrangian, there remains a possibility to find other solutions which lead
to the same contribution to the static energy $\widetilde{M}_V=0$. This, of
course, depends on the specific Lagrangian (more precisely the set of
parameters $\kappa _{n,m}$) but in general, it would lead to non-trivial
contributions from $G,H$ and $K$ in $M_{+}$ in (\ref{Mfghk}). Unfortunately,
it is easy to see that the same conclusions regarding the stability apply
for this case as well.

We note that even for arbitrary functions $G$, $H$ and $K,$ the quantities $%
a_V$ and $b_V$ are independent of the derivatives of $K.$ Indeed, recalling
for example the results of (\ref{avbv}) and (\ref{XiV}), we find that when
we consider only the static energy $\widetilde{M}_V$, it is minimized for%
$$
K=r\frac{H(G-1)^{\prime }-\ H^{\prime }(G-1)}{H^2+(G-1)^2}.
$$
This solution coincides with the field configuration in (\ref{gsin}) only in
the limits $\theta =0,\pi /2$ but these are just the relevant limits we are
interested in. In any case, rewriting $a_V$ and $b_V$ with the above
substitution leads to%
$$
\ a_V=\frac{J^2-1}{2r^2}\ \quad {\rm and\quad }a_Vb_V\ =\frac
1{4r^2}J^{\prime 2},
$$
where $J^2\equiv H^2+(G-1)^2$. The consequence of this last relation is that
it is possible to minimize the static energy for the gauge term with respect
to the function $J$ alone independently of the specific choice for $H$ or $G$%
. One could chose $H=0$ (or $G=1$) without loss of generality in which case $%
K=0$ and $J=G-1$ (or $J=H$). This symmetry is however explicitly broken by
the Lagrangian ${\cal L}_S$ as can be seen from $M_{+}$ in (\ref{Mfghk}).

\section{Other gauge invariant terms}

Since the gauge terms of the HGS Lagrangian is insufficient to guarantee
stability, it is also customary to add some sort of stabilizing Skyrme-like
term to ${\cal L}_{tot}$, preferably a ``gauge invariant'' Skyrme term of
the form \cite{Ezawa86,Klinkhamer86}:
\begin{equation}
\label{LGSkyrme}{\cal L}_{GSkyrme}={\frac 1{32e^2}} {\rm Tr\ } f_{\mu \nu
}f^{\mu \nu }+{\frac 1{32c^2}} {\rm Tr\ } d_{\mu \nu }d^{\mu \nu }
\end{equation}
where%
$$
\begin{array}{ccc}
f_{\mu \nu }= & R\left[ L^{\dag }D_\mu L-R^{\dag }D_\mu R,L^{\dag }D_\nu
L-R^{\dag }D_\nu R\right] R^{\dagger } & =\left[ L_\mu ,L_\nu \right] \\
d_{\mu \nu }= & R\left[ L^{\dag }D_\mu L+R^{\dag }D_\mu R,L^{\dag }D_\nu
L+R^{\dag }D_\nu R\right] R^{\dagger } &
\end{array}
{}.
$$
The first term is the usual Skyrme term. At this order (four derivatives),
these are the only contributions that lead to a positive Hamiltonian which
is second order in time derivatives.

It is most likely that a complete effective Lagrangian will involve
all-order terms. In that respect, the generalization of the previous
calculations to other gauge-invariant all-order terms constructed out of
powers of $d_{\mu \nu }$ is straightforward. The same constructions with $%
f_{\mu \nu }$ were considered in refs. \cite{Marleau89-10}. So we consider
an all-order Lagrangian $\widetilde{\cal L}_{+}$ of the form (\ref
{Traceszeta}) with $\zeta _{\mu \nu }\rightarrow $ $d_{\mu \nu }$. We may
attempt to construct a general Lagrangian or some special class. Since the
commutators $d_{\mu \nu }$ can be written as in (\ref{com}), the result will
inevitably be cast in a form like (\ref{Mstatic}) or even (\ref{Xi}) with $%
a_{+}$ and $b_{+}$ defined as
\begin{equation}
\label{afbfd}\ a_{+}=\frac 2{r^2}\left[ (G-1+\cos F)^2+H^2\ \right] \quad
{\rm and\quad }b_{+}=\frac 2{r^2}K^2
\end{equation}
We note that no derivatives of $F,G,H$ and $K$ are involved here at any
orders. Furthermore, for a gauge field configuration of the form (\ref{gsin}%
), the functions $a_{+}$ and $b_{+}$ become%
$$
\ a_{+}=\frac 2{r^2}\left[ g^2\sin {}^2\theta -2g\sin {}^2\theta (1-\cos
F)+(1-\cos F)^2\ \right] \ \quad {\rm and\quad }b_{+}=0.
$$
Then $\widetilde{\cal L}_{+}$ contribution to the static energy is
\begin{equation}
\label{Mplus}\widetilde{M}_{+}\equiv -\int d^3r\ \widetilde{\cal L}_{+}=4\pi
\int r^2dr\sum_{n=1}^\infty \kappa _na_{+}^n.
\end{equation}
Applying Derrick's test for stability under scale transformation could lead
to a stable soliton since there are terms of order ${\cal O}(r^{-2n})$ that
are potentially enough to stabilize the soliton. There is, for example,
absolute stability when all $\kappa _n>0$.

We conclude with a few remarks. The results presented here are only a
limited set of all-orders terms that could contribute to an effective
Lagrangian describing the hidden gauge symmetry and the scalar sector of the
electroweak theory. The analysis is limited to objects which have the
structure of powers of a commutator (any type of commutator, not only $%
f_{\mu \nu }$, $d_{\mu \nu }$ and $F_{\mu \nu }^V$). Mixed terms ({\it e.g. $%
{\rm Tr\ } (f_{\mu \nu }d^{\mu \lambda }\ldots )$) }were not considered
here. The generalization of our calculations to such terms is
straightforward but in absence of a specific model their interest remains
academic. Despite these limitations, the Lagrangians considered here are
attractive since they involve manifestly gauge-invariant all-orders
contributions in a very general way. This led to a number of results
regarding the energy of a static solution (mass of the soliton). In general,
higher-order terms can induce weak skyrmions but gauge terms alone are not
sufficient to absolutely stabilize the soliton. Most of the conclusions
reported here apply to QCD skyrmions where the vector meson is generally
interpreted as the $\rho $-meson. Finally, it is clear that whether or not
weak skyrmions exist, the presence of higher-order terms in an effective
Lagrangian for the strongly interacting scalar sector of the electroweak
theory are interesting since they do affect a number of observables. The
low-energy theorems however remain unaffected by the addition of higher
order terms since they generally rely on the lowest order terms \cite
{Harada93}.

This research was supported by the Natural Science and Engineering Research
Council of Canada and by the Fonds pour la Formation de Chercheurs et l'Aide
\`a la Recherche du Qu\'ebec.


\newpage

\begin{thebibliography}{99}
\bibitem{Chanowitz88}  M.S. Chanowitz, Ann.~Rev.~Nucl.~Part.~Sci.~{\bf 38}
323 (1988).

\bibitem{Veltman77}  M. Veltman, Acta~Phys.~Pol.~{\bf B8} 475 (1977); T.
Appelquist and C. Bernard, Phys.~Rev.~{\bf D22} 200 (1980).

\bibitem{Skyrme61}  T.H.R. Skyrme, Proc.~R.~Soc.~London.~{\bf A2603} 127
(1961).

\bibitem{Farhi81}  E. Farhi and L. Susskind, Phys.~Rep.~{\bf 74} 277 (1981).

\bibitem{Gipson81}  J.M. Gipson and H.C. Tze, Nucl.~Phys.~{\bf B183} 524
(1981).

\bibitem{Bando85}  M. Bando, T. Kugo, S. Uehara, K. Yamawaki and T.
Yanagida, Phys.~Rev.~Lett.~{\bf 54} 1215 (1985); M. Bando, T. Kugo and K.
Yamawaki, Phys.~Rep.~{\bf 164} 217 (1988).

\bibitem{Casalbuoni85}  R. Casalbuoni, S. de Curtis, D. Dominici and R.
Gatto, Phys.~Lett.~{\bf B155} 95 1985 ; Nucl.~Phys.~{\bf B282} 235 (1987);
A. Dobado and M.J. Herrero, Nucl.~Phys.~{\bf B319} 491 (1989).

\bibitem{Ezawa86}  Z.F. Ezawa and T. Yanagida, Phys.~Rev.~{\bf D33} 247
(1986).

\bibitem{Marleau89-10}  L. Marleau, Phys.~Lett.~{\bf B235} 141 (1990).

\bibitem{Marleau90-13}  S. Dub\'e and L. Marleau, Phys.~Rev.~{\bf D41} 1606
(1990); L. Marleau, Phys.~Rev.~{\bf D43} 885 (1991); A.D. Jackson, C. Weiss
and A. Wirzba, Nucl.~Phys.~{\bf A529} 741 (1991); K. Gustafsson and D.O.
Riska, Two-phase structure of infinite order skyrmion, University of
Helsinki preprint HU-TFT-93-11, January 1993.

\bibitem{Marleau92-11}  L. Marleau, Hidden gauge symmetry and the weak
skyrmions, Proc. of Beyond the Standard Model, Ottawa, June 22-24 1992,
World Scientific, ed. S. Godfrey.

\bibitem{Marleau91-11}  L. Marleau, Phys.~Rev.~{\bf D45} 1776 (1991).

\bibitem{Klinkhamer86}  F.R. Klinkhamer, Z.~Phys.~{\bf C31} 623 (1986).

\bibitem{Harada93}  M. Harada, T. Kugo and K. Yamawaki, Proving the low
energy theorem of hidden local symmetry, Kyoto University preprint
DPNU-93-01 and KUNS-1178, March 1993.
\end{thebibliography}
\end{document}